\documentclass[preprint,12pt]{elsarticle}

\usepackage{amssymb}
\usepackage{amsmath}
\usepackage{subfig}
\usepackage{booktabs}

\journal{Applied mathematical modeling}

\begin{document}

\begin{frontmatter}



\title{FourNetFlows: An efficient model for steady airfoil flows prediction}


\author[inst1]{Yuanjun Dai}
\affiliation[inst1]{organization={Department of Mechanics and Engineering Science, College
of Engineering},
            addressline={Peking University}, 
            city={Beijing},
            postcode={100871},
            country={China}}

\author[inst1]{Yiran An}
\author[inst1]{Zhi Li}


\begin{abstract}
FourNetFlows, the abbreviation of \emph{Fourier} Neural \emph{Net}work for Airfoil \emph{Flows}, is an efficient model that provides quick and accurate predictions of steady airfoil flows. We choose the Fourier Neural Operator (FNO) as the backbone architecture and utilize \emph{OpenFOAM} to generate numerical solutions of airfoil flows for training. Our results indicate that FourNetFlows matches the accuracy of the Semi-Implicit Method for Pressure Linked Equations (SIMPLE) integrated with the Spalart-Allmaras turbulence model, one of the numerical algorithms. FourNetFlows is also used to predict flows around an oval whose shape is definitely different from samples in the training set. We note that both qualitative and quantitative results are consistent with the numerical results. Meanwhile, FourNetFlows solves thousands of solutions in seconds, orders of magnitude faster than the classical numerical method. Surprisingly, FourNetFlows achieves model flows with zero-shot super-resolution when it is trained under a lower resolution. And the inferring time is almost constant when the resolution of solutions is increasing. 
\end{abstract}

\begin{graphicalabstract}
\includegraphics[totalheight=4in]{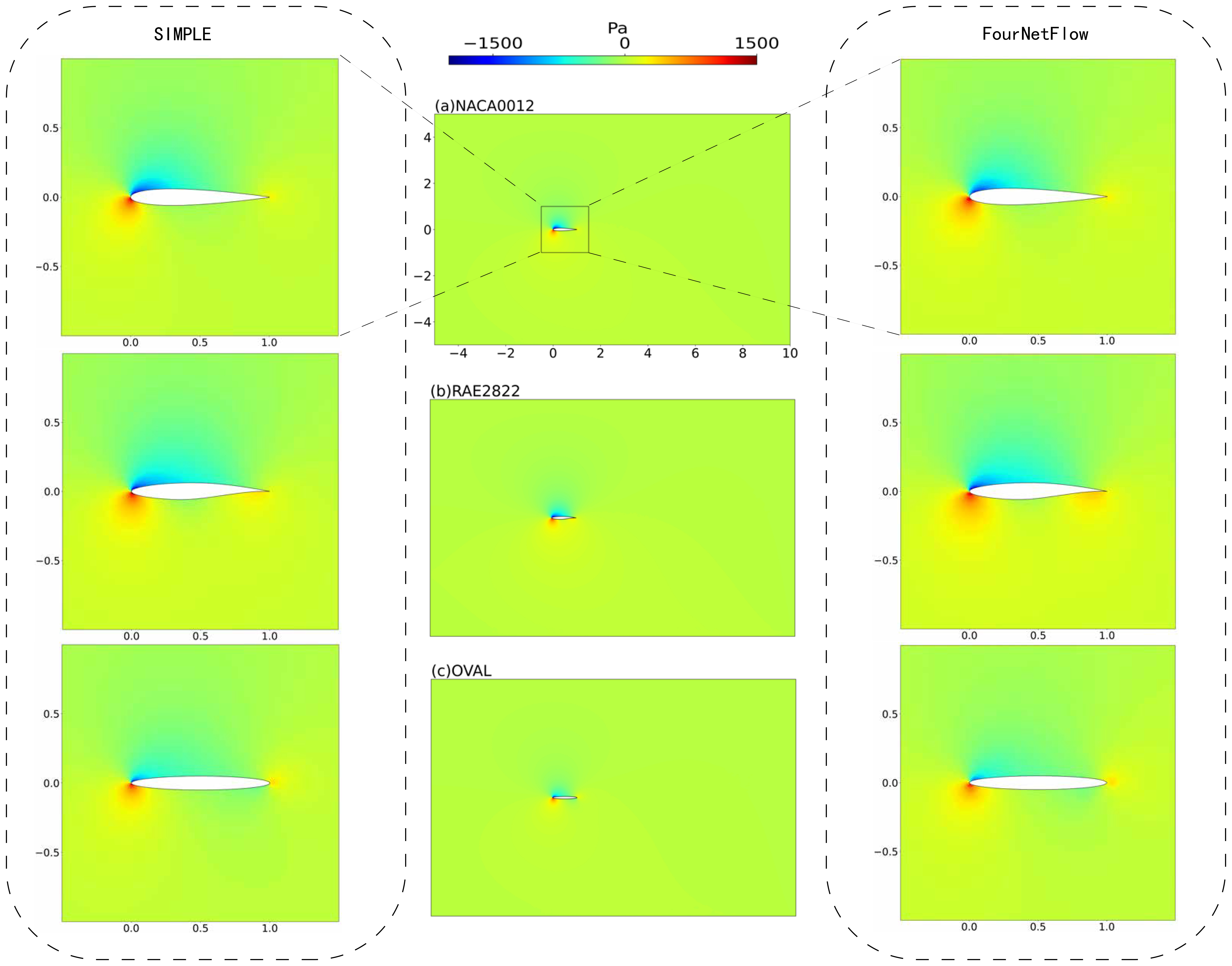}
\end{graphicalabstract}

\begin{highlights}
\item We propose an efficient model, FourNetFlows, mapped the boundary conditions to steady solution of airfoil flows. The accuracy of FourNetFlows is matched with the traditional method, running 4-5 orders of magnitude faster.
\item FourNetFlows has the ability of zero-shot super-resolution: trained on a lower resolution directly evaluated on a higher resolution. And the inferring time is almost constant when the resolution of solutions is increasing.
\item FourNetFlows is able to predict flows around an oval when it is trained by data of airfoil flows. Both qualitative and quantitative results are consistent with the numerical results.
\end{highlights}


\end{frontmatter}


\section{Introduction}
Numerical simulation of airfoil flows involves solving complex partial differential equations (PDEs) repeatedly for different values of boundary conditions, such as the geometry of airfoil profile, velocity and angle of attack. The numerical solution of PDEs is no mean feat, having led to a century of research so far. In light of recent successes in providing fast approximation of solutions by deep learning (DL) methods \cite{1,2,3,4}, it is becoming extremely difficult to ignore the role of DL methods in the field.
However, these works map between finite-dimensional spaces and therefore obtain some mesh-dependent models. This is often limitation for practical applications. There are some research proposed learning mesh-free, infinite-dimensional operators with neural networks \cite{5,6,7,8,9,10}. These neural operators are fast and accurate, even have the ability to transfer solutions between meshes at different resolution and need to be trained only once. But few studies take more practical boundary conditions into account. It is well-known that the solution of PDEs fully depends on boundary conditions for a steady problem.
Part of the aim of this paper is to develop an efficient neural operator for steady airfoil flows, considering the geometry of airfoil profile, velocity and angle of attack. 

\begin{figure}
  \centering
  \includegraphics[totalheight=4in]{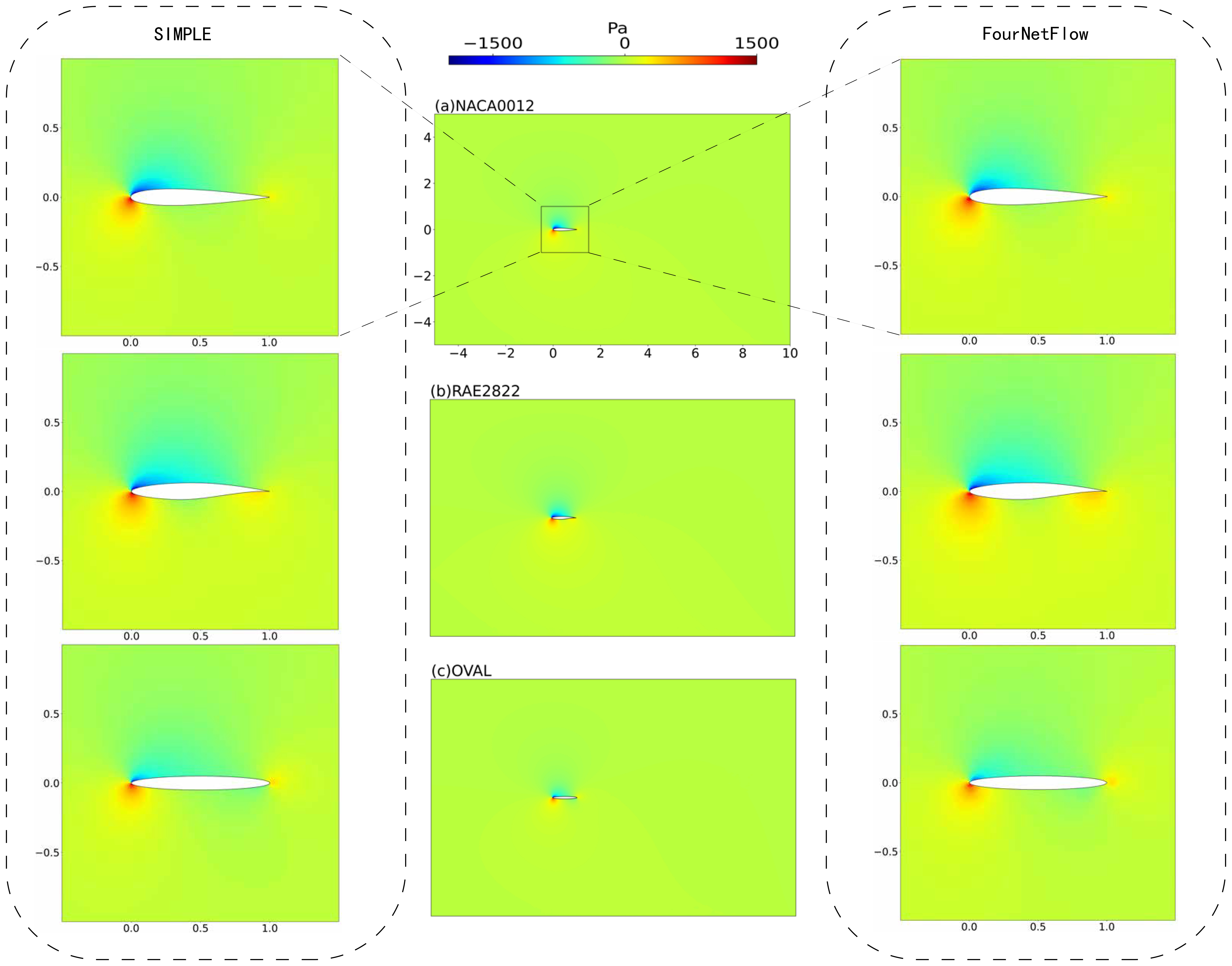}
  \caption{Illustrative example of pressure fields generated by FourNetFlows and SIMPLE at resolution of 1024 x 1024. Panel(a) shows the pressure field of the NACA0012, which has been widely used for validating a new algorithm due to the simplicity of the geometry. Panel(b) shows the model prediction of the pressure field of the RAE2822. It is a kind of typical supercritical airfoil. Panel(c) shows the pressure field of OVAL10.}
  \label{fig:fig1}
\end{figure}

Our model, FourNetFlows, is based on Fourier Neural Operator (FNO) \cite{10}, the state-of-the-art (SOTA) neural operator. We utilize \emph{simpleFoam}, a numerical solver based on Semi-Implicit Method for Pressure Linked Equations (SIMPLE), to generate ground truth data with the University of Illinois at Urbana-Champaign (UIUC) database. Figure \ref{fig:fig1} shows an illustrative entire pressure field of NACA0012, RAE2822, and OVAL10 (an oval with a ratio of 10 between the long and short axes). We highlight the key zone that are solved by \emph{SIMPLE} (\emph{simpleFoam} in \emph{OpenFOAM}) and accurately predicted by our model. The comparison qualitatively shows that FourNetFlows has the ability to predict physical fields. We quantitatively shows the accuracy of FourNetFlows is matched with the traditional method, running 4-5 orders of magnitude faster. Meanwhile, it can directly generate the flow field at resolution of $1024 \times 1024$ when trained by samples at $128\times128$. A potential problem is that we have to use Cartesian grid to describe computational domain due to the limitation of discrete Fourier transform (DFT). The Cartesian grid is capable of distinct the large scales of airfoil features, but ignore the smaller scales. Fortunately, the problem will be alleviated with finer grids, since the computational cost of FourNetFlows is almost constant when the resolution is increased.

\section{Related work}
\label{sec:related work}
Most of detailed mathematical models of physical phenomena are expressed naturally in PDEs.
Solving PDEs with machine learning (ML) algorithms has become an area of great research interest in computational fluid dynamics (CFD) \cite{11}. Since high resolution simulations or experiments are often too expensive and too slow. While learned models provide faster predictions, reducing turnaround time for workflows in engineering and science \cite{12,13,14,15,16}. We distinguish modeling flow problems by ML algorithms into two major branches: modeling flow kinematics through the extraction flow features and modeling flow dynamics through the adoption of various learning architectures. In the aspect of flow feature extraction, the most common approach include dimensionality reduction, clustering and classification. Extensive research has shown that these algorithms is promise for flow problems. Milano \& Koumoutsakos used a simple neural network to substitute for the standard proper orthogonal decomposition or principal component analysis (POD/PCA), reconstructing the near wall velocity field in a turbulent channel flow with wall pressure and shear\cite{17}. Kaiser et al. employed the k-means algorithm to discovery a low-dimensional representation of a high-dimensional phase space for the fluid mixing layer. With these clusters, Kaiser et al. utilized tractable Markov transition models to simulate how the flow evolves in time from one state to another \cite{18}. Colvert et al. investigated the classification of wake topology behind a pitching airfoil via taking local vorticity measurements as inputs of neural networks\cite{19}. Closer to the goals of our work, Li applied reduced-order modeling (ROM) and long short-term memory (LSTM)\cite{20}, a revolutionary model for speech recognition, to predict unsteady aerodynamics of NACA 6-series airfoil\cite{21}. In the aspect of modeling flow dynamics through various neural networks architecture, early examples include the use of neural networks to learn the solutions of ordinary and partial differential equations$^{[22\textrm{--}24]}$. The potential of these works has not been fully explored and in recent years there is further advances $^{[1\textrm{,}25]}$. We note also the possibility of using these methods to uncover latent variables and reduce the number of parametric studies often associated with PDEs. Zhang et al. developed a convolutional neural network (CNN), the most famous architecture in computer visions, to infer the lift coefficient of airfoils\cite{15}. Thuerey et al. proposed a DL-based method for RANS simulation of airfoil flows via UNET, a neural networks architecture in image segmentation\cite{26}. Wan et al. and Vlachas et al. used LSTM to model dynamical systems and for data driven predictions of extreme events\cite{27}. Generative adversarial networks (GANS), proposed by Goodfellow et al.\cite{28} for generating images, are also being used in fluid dynamics\cite{29}. We note that even in these quite effective works, they directly took some network architecture as backbones by analogizing the fluid problem to problems in other domains. This clearly contradicts the \emph{no free lunch theorems}. Therefore, we propose FourNetFlows based on FNO, which is designed for modeling flows.
Our contributions can be broken down into three main parts.
\begin{itemize}
\item We take boundary conditions as inputs of FourNetFlows instead of initial conditions and locations which are the inputs of FNO. This make our model are more practical in investigation of airfoil flows.
\item A convolutional layer is substituted for the first linear layer in FNO, which improve our model accuracy. 
\item Solution generalization from airfoil flows to flows around an oval.
\end{itemize}

This paper is structured as follows. In Section \ref{sec:related work}, we introduce some works in airfoil flows predictions by DL methods and briefly outline differences. In Section \ref{sec:method}, the FourNetFlows and why it can zero-shot super-resolution is discussed in detail. In Section \ref{sec:experiments}, we evaluate our model accuracy and computational cost. Finally, conclusions are provided in Section \ref{sec:conclusion}. More details about data generation, data pre-processing and output post-processing are list in Appendix. 

\section{Method}
\label{sec:method}
\begin{figure}
  \centering
  \includegraphics[totalheight=6in]{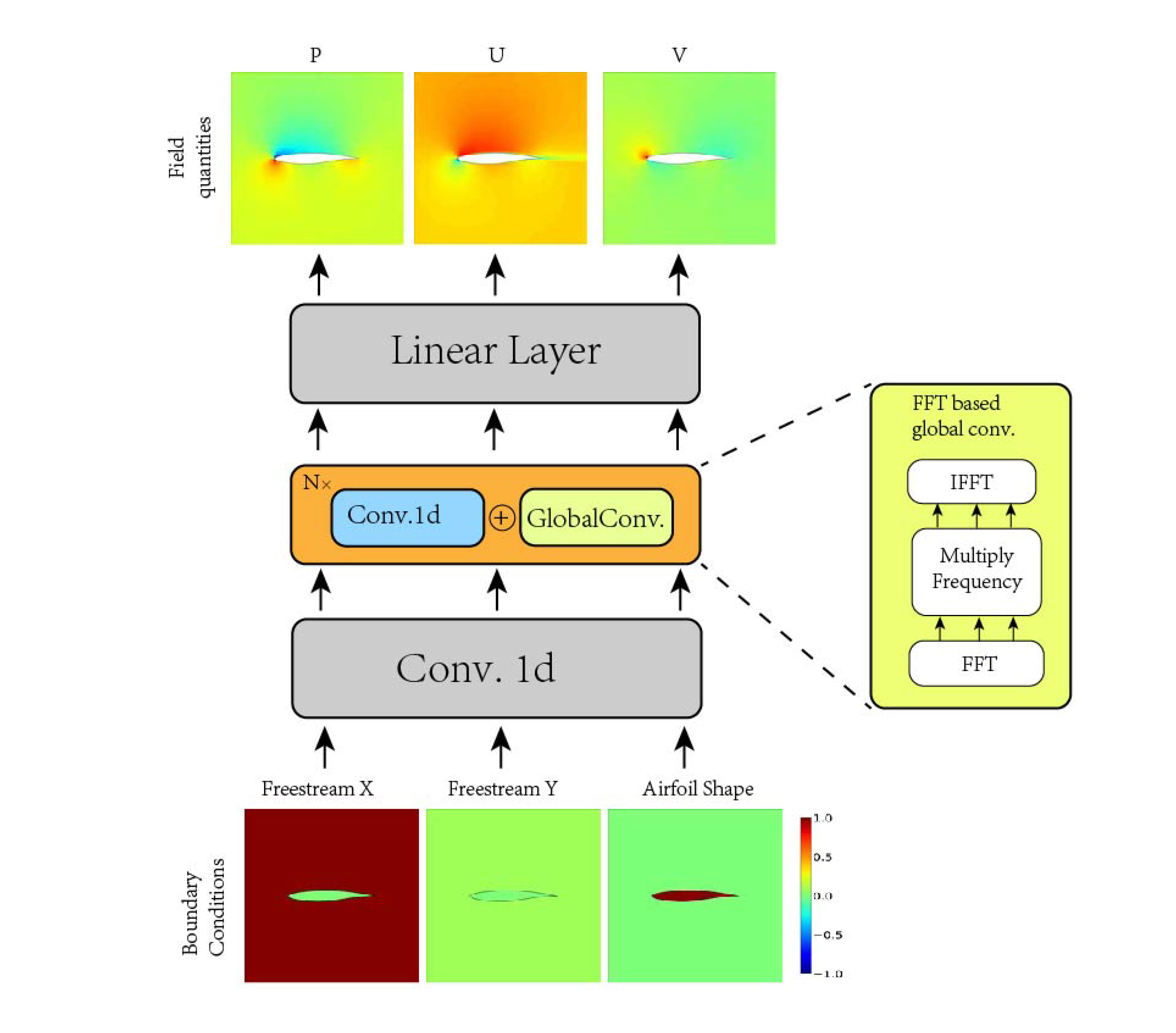}
  \caption{The full architecture of the neural networks. Start from boundary conditions, which include airfoil shape, freestream velocity in x-direction and y-direction. First, map the three boundary conditions to a function by a local convolution. Then, apply N times global convolutions and activation functions. Finally, decode the target physical field.} 
  \label{fig:fig2}
\end{figure}
To my best knowledge, there is not yet a publicly available and reliable dataset of steady airfoil flows. Fortunately, UIUC provides an airfoil dataset so that it is not hard to generate numerical data for training. Details of data generation and data processing see \ref{appendix a}. 

In this work, we focus on two important and challenging task, namely, (1) accurately solving the velocity and pressure distribution of flows around airfoils and (2) super-resolution, which means transferring the solution between meshes at different resolution. There are a few reasons for our focus on these tasks. First, accurately predicting the velocity and pressure field is the key to evaluate the aerodynamic performance of airfoils. It is the basic requirement for a model whose goal is to function in practical engineering. Second task is super-resolution, a weak point for classical numerical methods. Numerical methods always impose a trade-off on the resolution of solutions: coarse grids are fast but less accurate, fine grids are accurate. However, this happens to be an area where deep learning (DL) models can potentially show great promise.

\subsection{Architecture}
To map accurate solution from boundary conditions, we choose the FNO as the backbone. This particular neural network architecture is SOTA in modeling challenging PDE systems with periodic boundary conditions. The FNO model is unique in that it not only apply a local convolution on inputs, but also a continuous global convolution that implemented efficiently in the Fourier domains with fourier transform. That allows modeling dependencies across spatial dimensions flexibly. With such a design, FNO is well-suited to zero-shot super-resolution. In the original FNO formulation, the operator learning approach is designed for flows with periodic boundary conditions. A FNO variant with characteristics-based boundary conditions have been proposed by us.

At first, we briefly introduce the basic idea of neural operators for solving PDEs. When we consider PDEs in the following form on the unit box:
\begin{equation}
\begin{aligned}
    (\mathcal{L}_au)(x) &= f(x),~~~~~x\in D\\
    u(x) &= 0, ~~~~~x \in \partial D
\end{aligned}
\end{equation}
Under fairly general conditions on $\mathcal{L}_a \cdot = -\mathbf{div}(a\nabla \cdot)$, the Green's function $G:D\times D\rightarrow \mathcal{R}$ as the unique solution to the problem can be defined.
\begin{equation}
    \mathcal{L}_aG(x,\cdot) = \delta_x 
\end{equation}
The solution can then be represented as:
\begin{equation}
    u(x) = \int_D G_a(x,y)f(y) dy
\end{equation}
Due to the cost of evaluating integral operators, it is not easy to model the formulation via a neural operator. Through the Fourier transform, the FNO alleviates this issue. While We refer the reader to the original paper for more details, we talk more about differences between FourNetFlows and FNO. The major difference is inputs of models. The inputs of FNO are the solution of the coefficient function and locations when the FNO is used to solve the 2-D Darcy Flow equation. It is certainly feasible for the PDEs. Since the mapping from $a$ to $u$ is unique under the limitation of zero boundary on $\partial D$. As for airfoil flows, the boundary conditions are more complex. The physical fields are fully determined by the velocity and the shape of airfoils. We take the following form as the governing equations:
\begin{equation}
\begin{aligned}
    \textbf{u}(x)&=\mathcal{N}_1(\textbf{u}_\infty,s),~~~~~x\in D\\
    p(x)&=\mathcal{N}_2(\textbf{u}_\infty,s),~~~~~x\in D\\
\end{aligned}
\end{equation}
where $\textbf{U}_\infty$ is the freestream velocity, $s$ represents the geometry of airfoils profile. $\mathcal{N}_1$ and $\mathcal{N}_1$ are non-linear operators, equivalent to the continuous equation and the Navier-Stokes equation.
The inputs of FourNetFlows are split into three parts, namely, the airfoil shape, freestream velocity in x-direction and y-direction. We discard locations as one of inputs since the discrete Fourier transform always considers the sampling points to be uniformly distributed. The formula of DFT as follows:
\begin{equation}
    X[k] = \sum_{n=0}^{N-1}x[n]e^{j(\frac{2 \pi}{N})kn},~~~~~k\in[0,N-1]
\end{equation}
Another difference between FourNetFlows and FNO is the first layer of models. The FNO projects inputs to a higher dimension space by the layer linear. The FourNetFlows utilizes the layer, a Conv.1d, to map boundary conditions to a tensor with mutiple channels. This change reduces the cost function by about 20\%, from $3.4 e^{-3}$ to $2.9 e^{-3}$. The rest of neural networks are the same. Figure \ref{fig:fig2} shows the full architecture and the flow of computation in our model. The computational flow of our model as follows:
\paragraph{Step 1} Transform boundary conditions to a feature tensor.
\begin{equation}
    y_{m,n,c}^0 = \sigma(W x_{m,n,3})
\end{equation}
where $m,n$ represents the number of grids in x-direction and y-direction, $W$ represents the local convolution kernel, $c$ denotes the number of channels.
\paragraph{Step 2} Applied N times local and global convolution on the feature tensor.
\begin{equation}
    y_{m,n,c}^i = \sigma(W y_{m,n,c}^{i-1}+\mathcal{F}^{-1}(R_{\phi}\cdot\mathcal{F}(y_{m,n,c}^{i-1})))
\end{equation}
where $\mathcal{F}$ denote the Fourier transform and $\mathcal{F}^{-1}$ its inverse.
\paragraph{Step 3} Deconde and obtain target physical fields.
\begin{equation}
    T_{output} = B^T y_{m,n,c}^N + bias
\end{equation}
where $B^T$ represents the weights of the linear layer, $T_{output}$ is the output tensor.
\subsection{Zero-shot Super-resolution}
The quest for high-resolution flow data has been one of the major pursuits in traditional solvers. The common approach is that reconstruct a finer flow field by interpolation methods based on a coarse flow field. Instead of doing this, It is better to generate the high-resolution physical field directly. The FourNetFlows shares the same learned network parameters irrespective of the discretization used on the input and output spaces since the nature of Fourier transform. So It can do zero-shot super-resolution: trained on a lower resolution directly evaluated on a higher resolution, as shown in Figure \ref{fig:fig1}

\section{Experiments}
\label{sec:experiments}
We evaluate the modeling Navier-Stokes equation capacities of FourNetFlows by three different type airfoil,
namely, NACA0012, RAE2822 and OVAL10. We show the flow field predicted by the FourNetFlows are in agreement with those solved by SIMPLE. After post-processing, the aerodynamic coefficients also remain consistent with experimental results. When considering the computational cost, FourNetFlows performs very favourably. On a $128 \times 128$ grid, the FourNetFlows has an inference time only 1 second compared to a minute of the SIMPLE used to solve Navier-Stokes equation. In particular, FourNetFlows's computation speed is even more impressive when computing a bulk of cases, with it taking only about 6 seconds to process 1280 cases.
\paragraph{Boundary Conditions} We consider the 2D airfoil flow as a viscous and incompressible problem. It is a reasonale assumption when the Mach number is less then 0.3. With the assumption, we don't need a input tensor of pressure. The freestream velocity $U=$ 50 m/s and the angle of attack $\alpha =5^{\circ}$. The Reynolds number $Re \approx 5$ millions.
\paragraph{SIMPLE Setup} The numerical simulation make use of the widely used Spalart-Allmaras turbulence model and solutions are calculated with the open source code \emph{OpenFOAM}. The setup is the same like data generation. More details see \ref{appendix a}. 
\paragraph{FourNetFlows Setup} There are only 5 hyperparameters in our model. $N$ is the number of the Fourier layer. $c$ is the number of feature tensor's channels. $(h, w)$ is the shape of $R_{\phi}$ in Fourier space and $k_{\text{max}}$ is the maximal numbel of modes at which the Fourier series are truncated. In the trained model, $N = 4$, $c=128$, $(h, w) = (32, 32)$, and $k_{\text{max}} = 12 
$.
\subsection{Physical Fields}
Figure \ref{fig:fig1} qualitatively shows the predict ability of our FourNetFlows model on forecasting these pressure fields at a resolution of $1024\times1024$ when the model is trained with a resolution of $128\times128$. These pressure fields are obtained by mutiplying the FourNetFlows output tensor by the maximum pressure of the training set i.e., $P = T_{\text{output}}\times P_{\text{max}}$. Starting from airfoil shapes from the out-of-sample test dataset, the model was allowed to run freely along the computational flow (Figure \ref{fig:fig2}) and output $T_{\text{output}}$. Figure \ref{fig:fig1} (a)-(c) show the pressure field of different airfoils predicted by the model (right-panel) and the corresponding pressure field solved by SIMPLE. We note that the FourNetFlows model is able to predict the pressure around different type of airfoils in advance with remarkable fidelity with correct fine-scale features. For example, for RAE2822, a supercritical airfoil, the trailing edge pressure distribution is completely different from that of a symmetrical airfoil. A high pressure area should appears on the lower trailing edge to compensate for the lack of lift. Our model predicts the high pressure area perfectly. When the airfoil shape is generalized to an oval, our model predictions also remain consistent with the numerical solution. The FourNetFlows model appears to be able to forecast physical field for flow around solid whose shape different away from an airfoil.

\begin{figure}
  \centering
  \includegraphics[totalheight=7in]{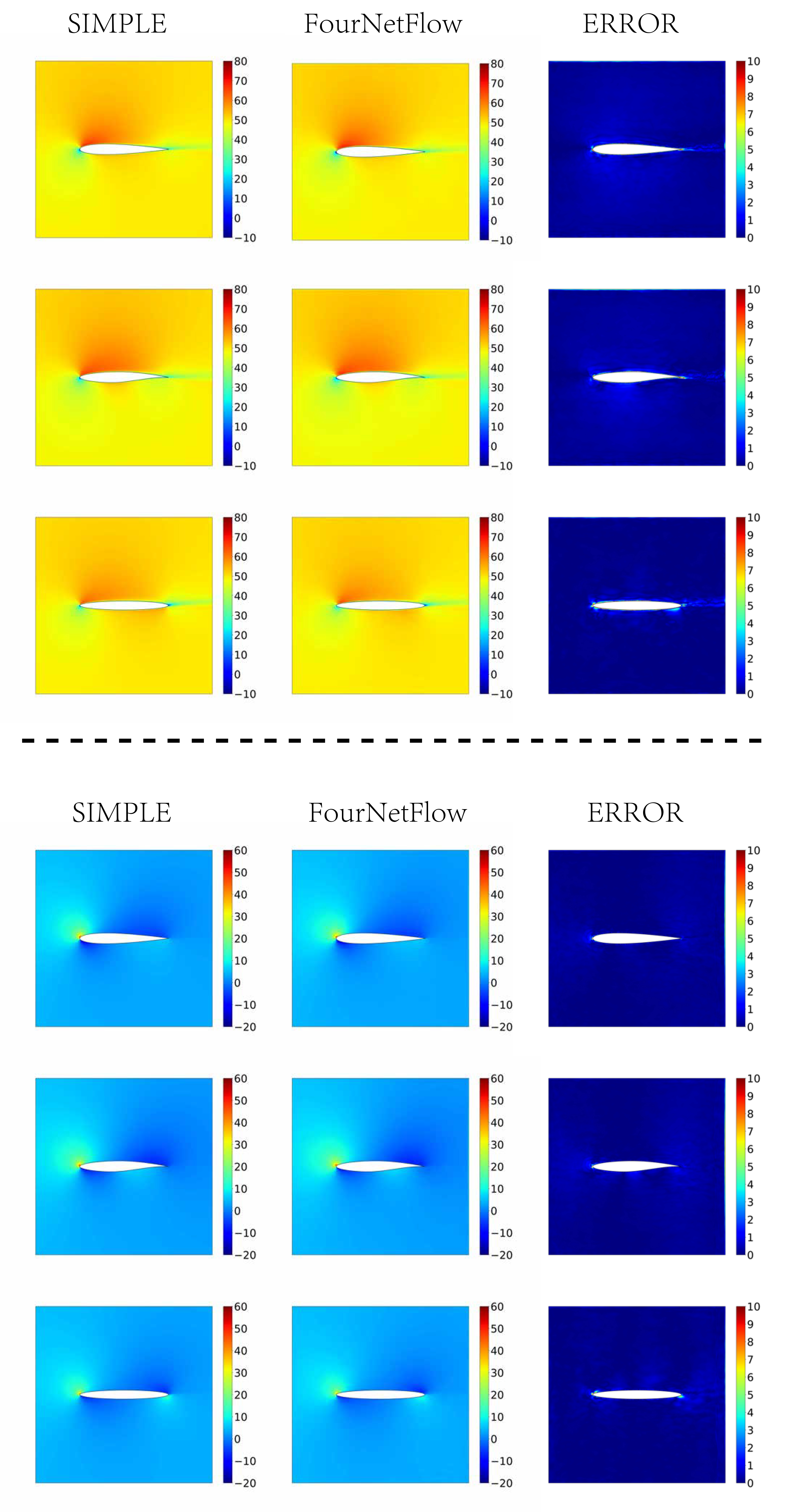}
  \caption{Illustration of velocity fields forecast using the FourNetFlows. The top panel shows the comparison of the component of velocity in x-direction solved by SIMPLE and that predicted by the FourNetFlows. The bottom panel shows the comparison of the component of velocity in y-direction solved by SIMPLE and that predicted by the FourNetFlows.} 
  \label{fig:fig3}
\end{figure}

The top and bottom panel of Fig \ref{fig:fig3} respectively provide a qualitatively visualization of the velocity component in x-direction and y-direction, highlighting the ability of the FourNetFlows model to forecast the velocity field with remarkable accuracy. The column of \emph{ERROR} shows the absolute difference between velocity predicted by the FourNetFlows and SIMPLE. It shows the major difference of the velocity predicted by the two methods are located around airfoils. 
This could be due to two reasons. On the one hand, the velocity field of the SIMPLE is interpolated from the original grid and the interpolation may have introduced errors. On the other hand, the gradient of velocity in the area near walls is relatively large. We observe that for the oval case the major differences occur near the trailing edge. For flow around an oval, it is not well-suited to governed by the steady-state Navier-Stokes equation. This problem is more serious in cases where the ratio of long to short axis is smaller. And the shape of the oval is far away from shapes of airfoil in training set. Generalization has always been a challenge for neural networks. Despite it is challenging, we note that the model shows excellent skill in generalization, which can have significant meaning to develop a more general solver based the FourNetFlows. From the results, we believe the model really learns the nature of flow around a body.
\subsection{Quantitative Skill of FourNetFlows}
We post-process the flow fields of these cases to quantitatively evaluate the predictive ability of the model. For each airfoil, we estimate the coefficient of pressure, $C_p$, under the boundary condition of $\alpha = 5^{\circ}$ and $U = 50$ m/s. In practical engineering, the physical quantities that designers care about actually integral quantities, such as the coefficient of lift, $C_l$, the coefficient of drag, $C_l$, and the coefficient of pitching moment, $C_m$. These quantities are crucial to the aerodynamic performance of airfoils. Therefore, we compares $C_l$, $C_d$, $C_m$ predicted by FourNerFlow with these quantities solved by the SIMPLE under different angle of attack. The definition of $C_p$, $C_l$, $C_d$, $C_m$ and more details about post-process see \ref{appendix a}. 

\begin{figure}
\centering
\subfloat[]{%
\includegraphics[totalheight=1.2in]{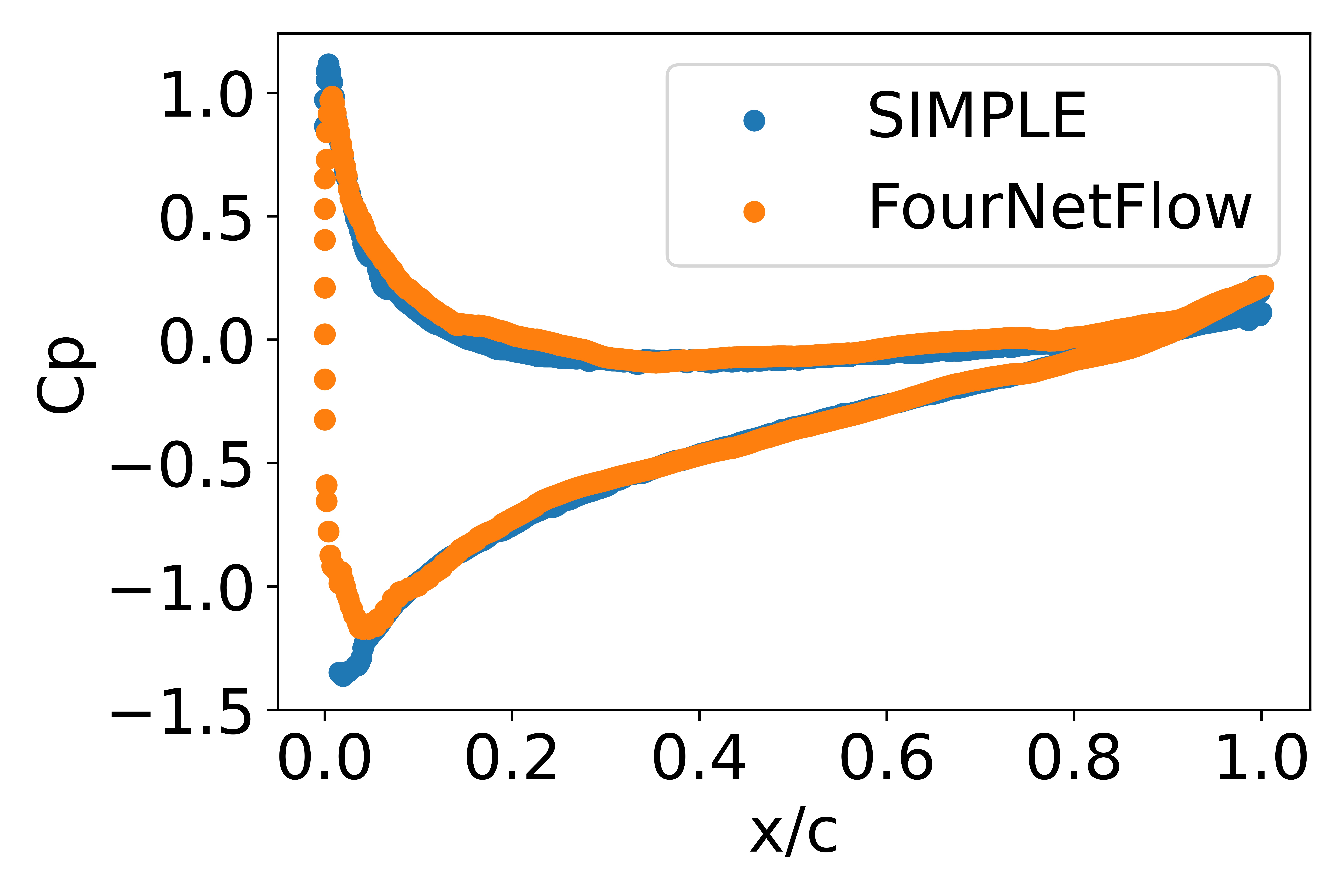}}
\subfloat[]{%
\includegraphics[totalheight=1.2in]{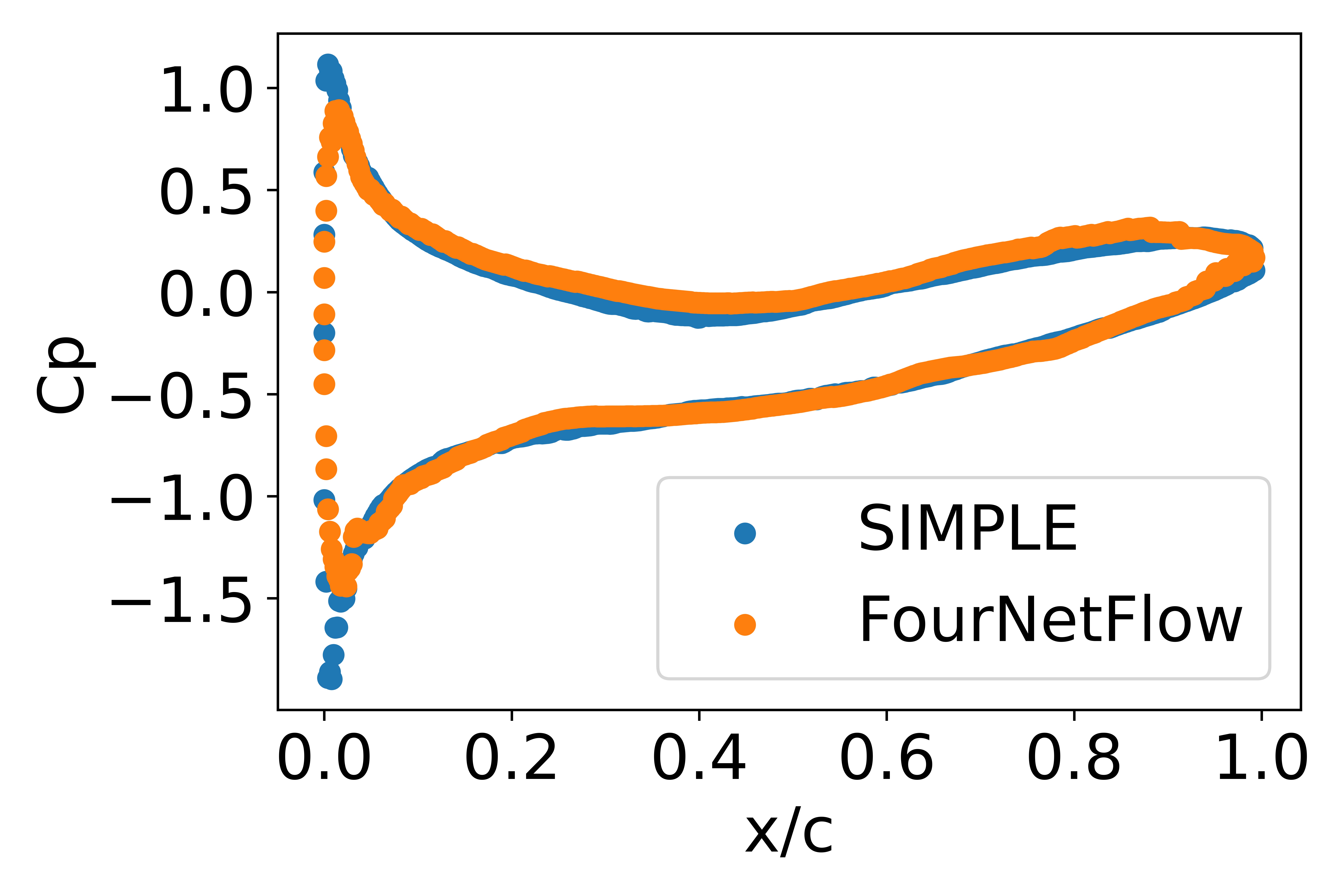}}
\subfloat[]{%
\includegraphics[totalheight=1.2in]{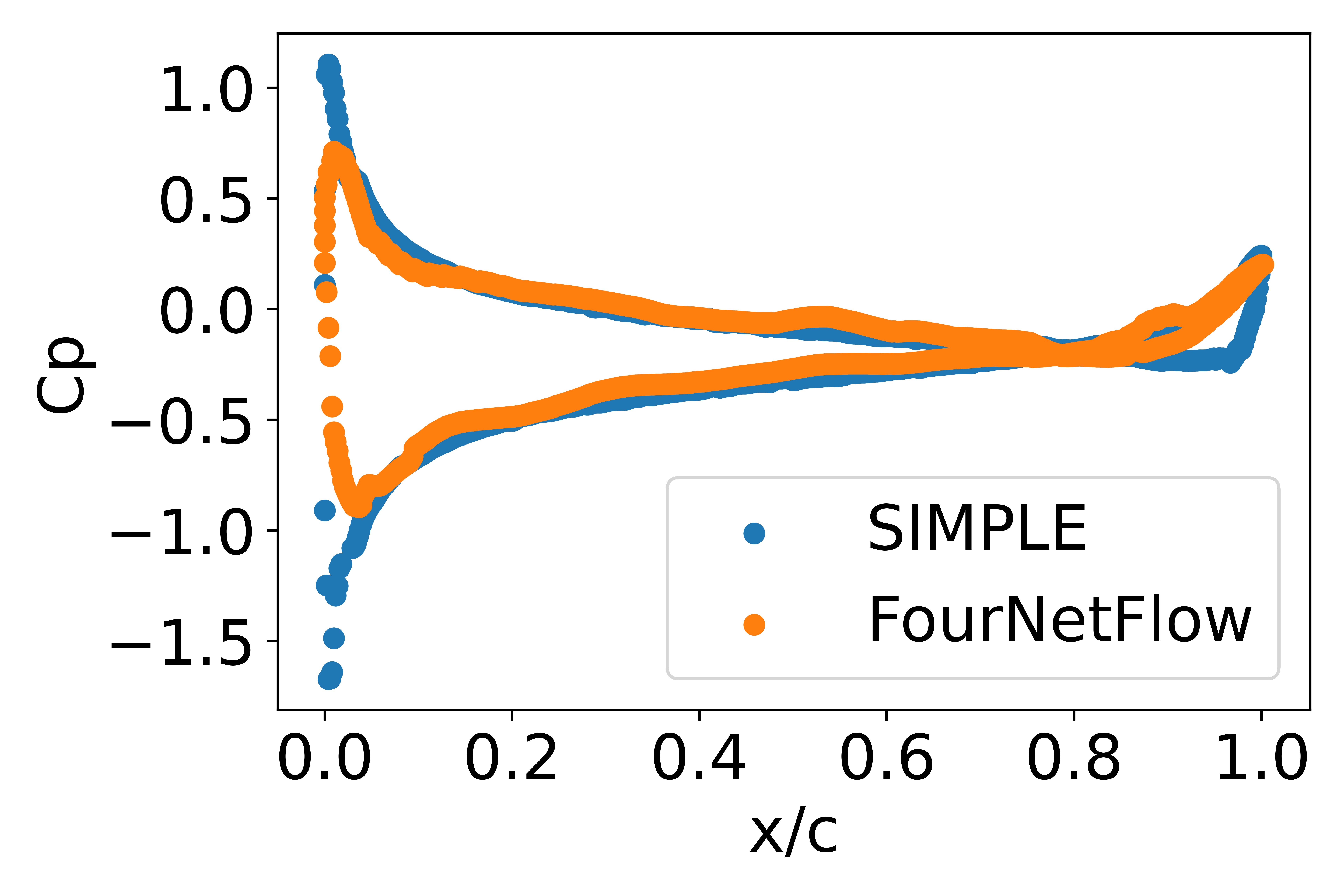}}
\caption{Comparison of $C_p$ predicted by two methods. Panel(a)-(c) respectively shows the $C_p$ of NACA0012, RAE2822 and OVAL10.}
\label{fig:fig4}
\end{figure}

We show our comparison $C_p$ for in Figure \ref{fig:fig4}. We find that the FourNetFlows achieve similar $C_p$ in most area of airfoils. Even in the case of the oval, our model performs well. The major difference of two methods still located at the leading edge and the trailing edge. The reason for this is that the pressure changes steeply here. Our model always tends to give a smoother solution. In some of works using neural networks to forecast $C_p$ \cite{21}, predictions of their model match the results of CFD better than ours. But our work is fundamentally different from these works. They directly take the $C_p$ as the learning target, whereas we obtain the $C_p$ from post-processing the flow field predicted by our model. Comparison of $C_p$ is aim to quantify how well our model predicts the physical field, not just for the $C_p$.

Figure 5 (a)-(i) shows these integral quantities for the FourNetFlows model predicts and the corresponding matched the SIMPLE. A good agreement is present for both two methods. 
The $C_l$ is essentially linearly distributed along angles of attack, which is in line with the characteristics at small angles of attack. The errors of $C_d$ and $C_m$ are large when compared with the errors of $C_l$ since $C_d$ and $C_m$ are small quantities relative to $C_l$. Our integration method differs from that used in \emph{OpenFOAM}. The errors introduced by integration may have little effect on $C_l$ but a greate effect on these two small quantities. We specify the integration method we used in post-processing in the \ref{appendix a}. 

In general, the FourNetFlows are very competitive with traditional solvers, both in terms of comparison of integral quantities and details of flow field. 

\begin{figure}
\centering
\subfloat[]{%
\includegraphics[totalheight=1.2in]{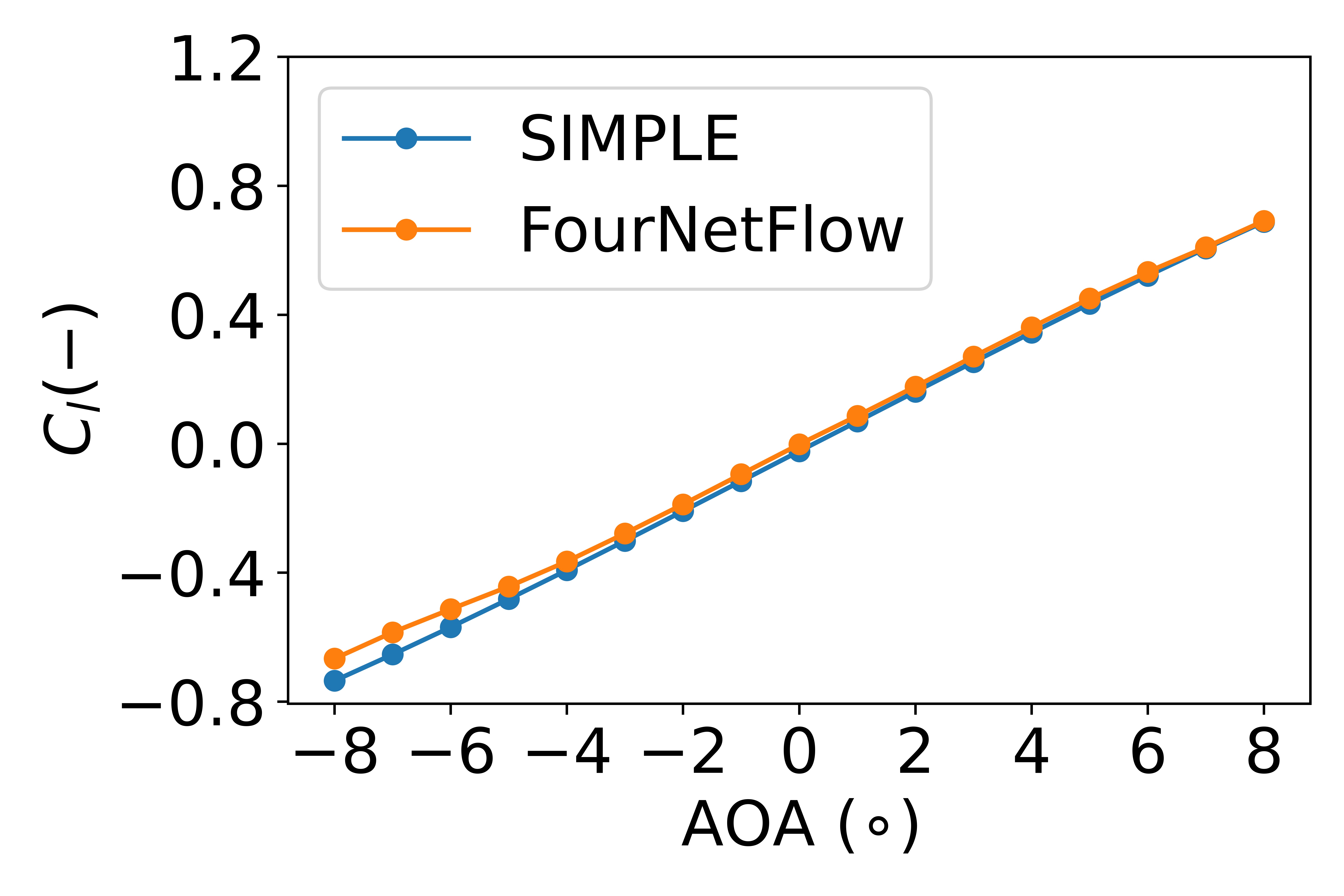}}
\subfloat[]{%
\includegraphics[totalheight=1.2in]{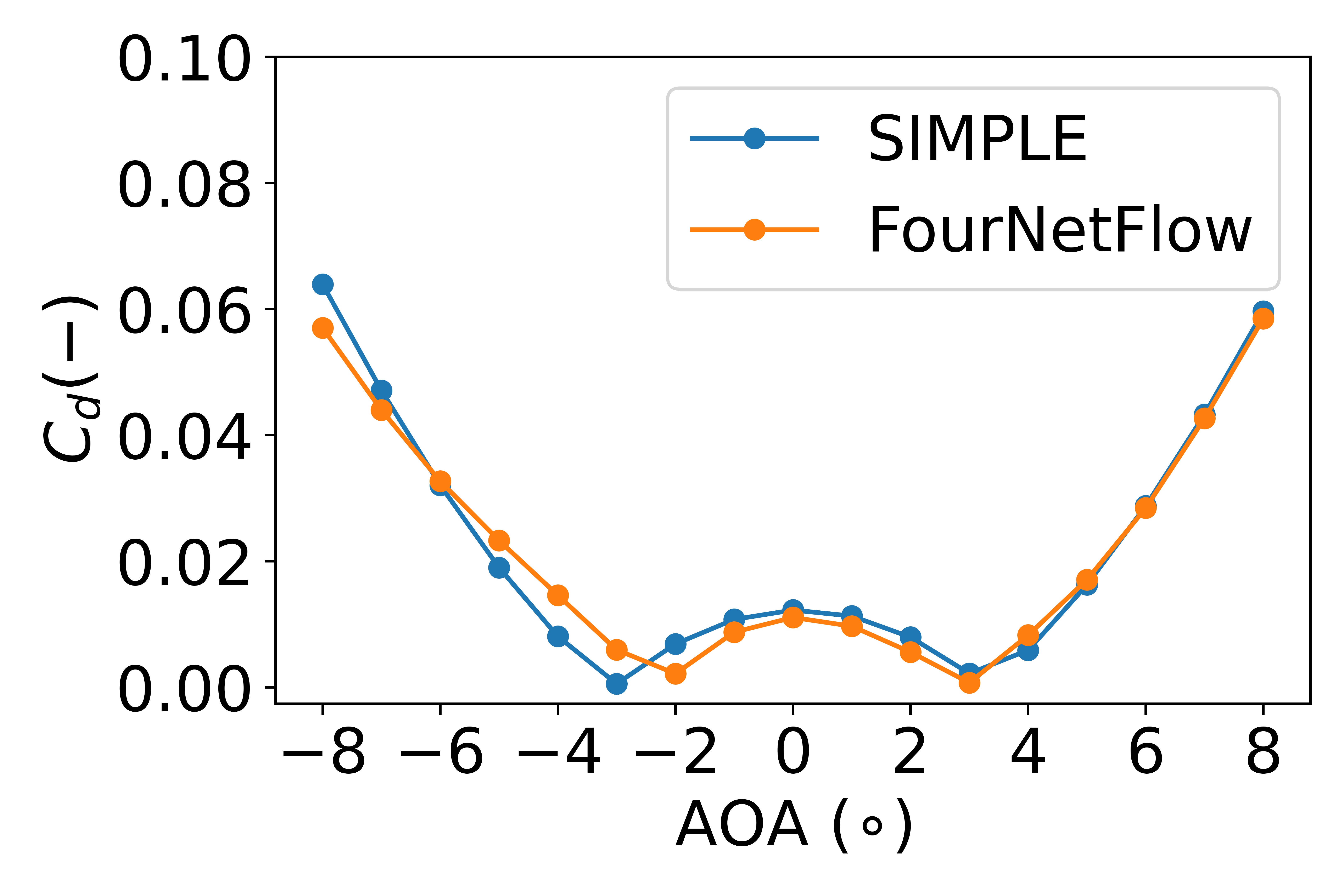}}
\subfloat[]{%
\includegraphics[totalheight=1.2in]{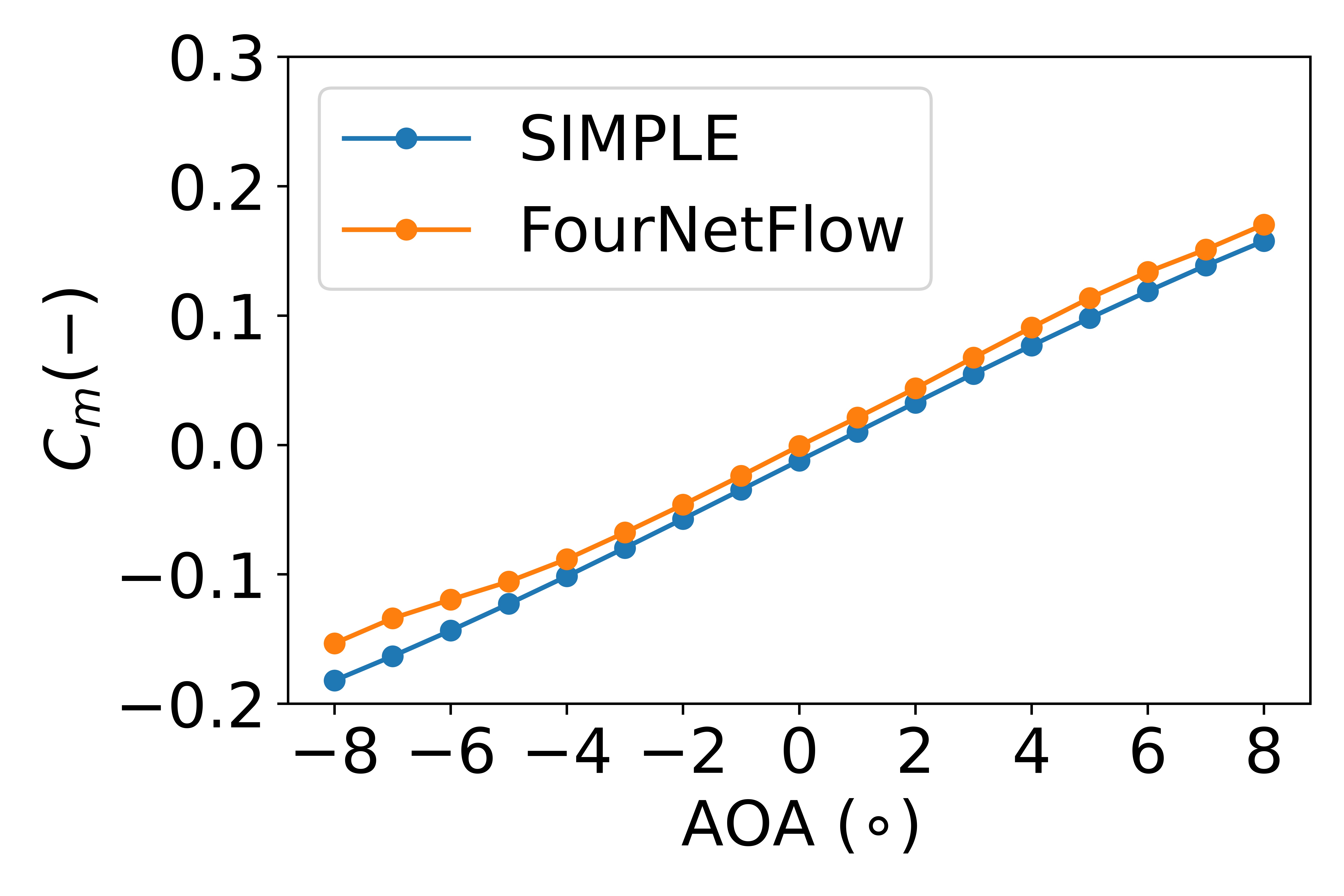}}
\quad
\subfloat[]{%
\includegraphics[totalheight=1.2in]{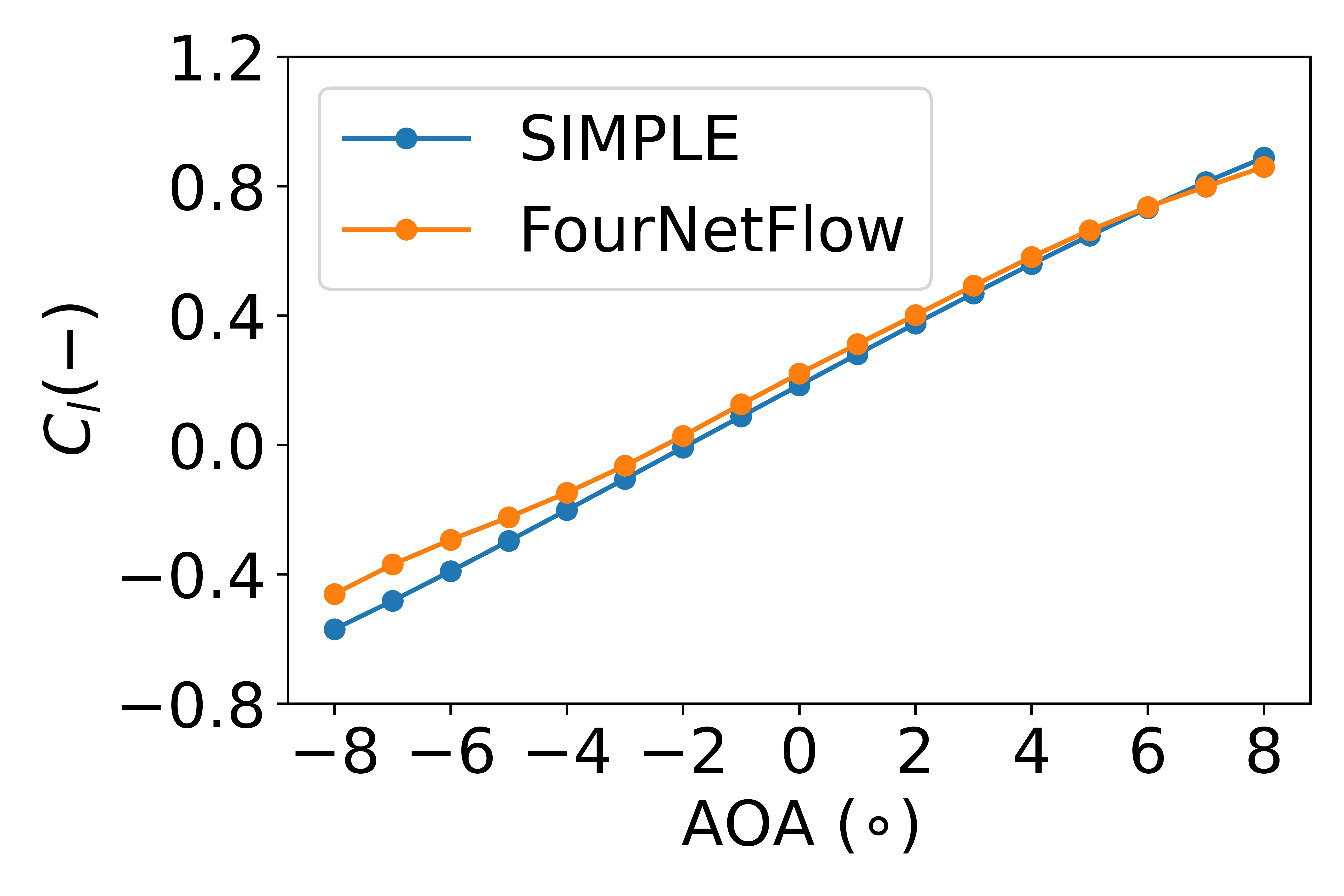}}
\subfloat[]{%
\includegraphics[totalheight=1.2in]{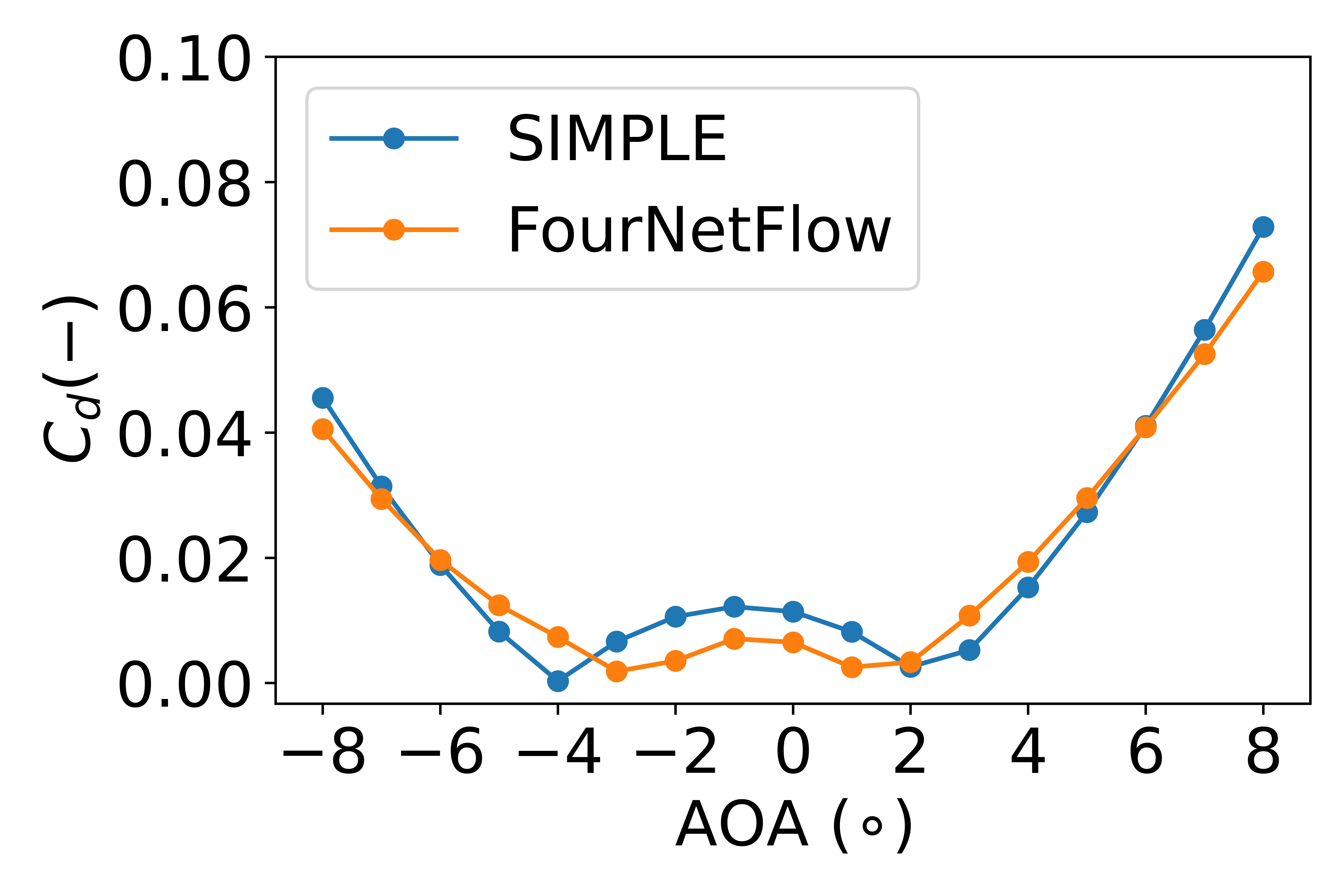}}
\subfloat[]{%
\includegraphics[totalheight=1.2in]{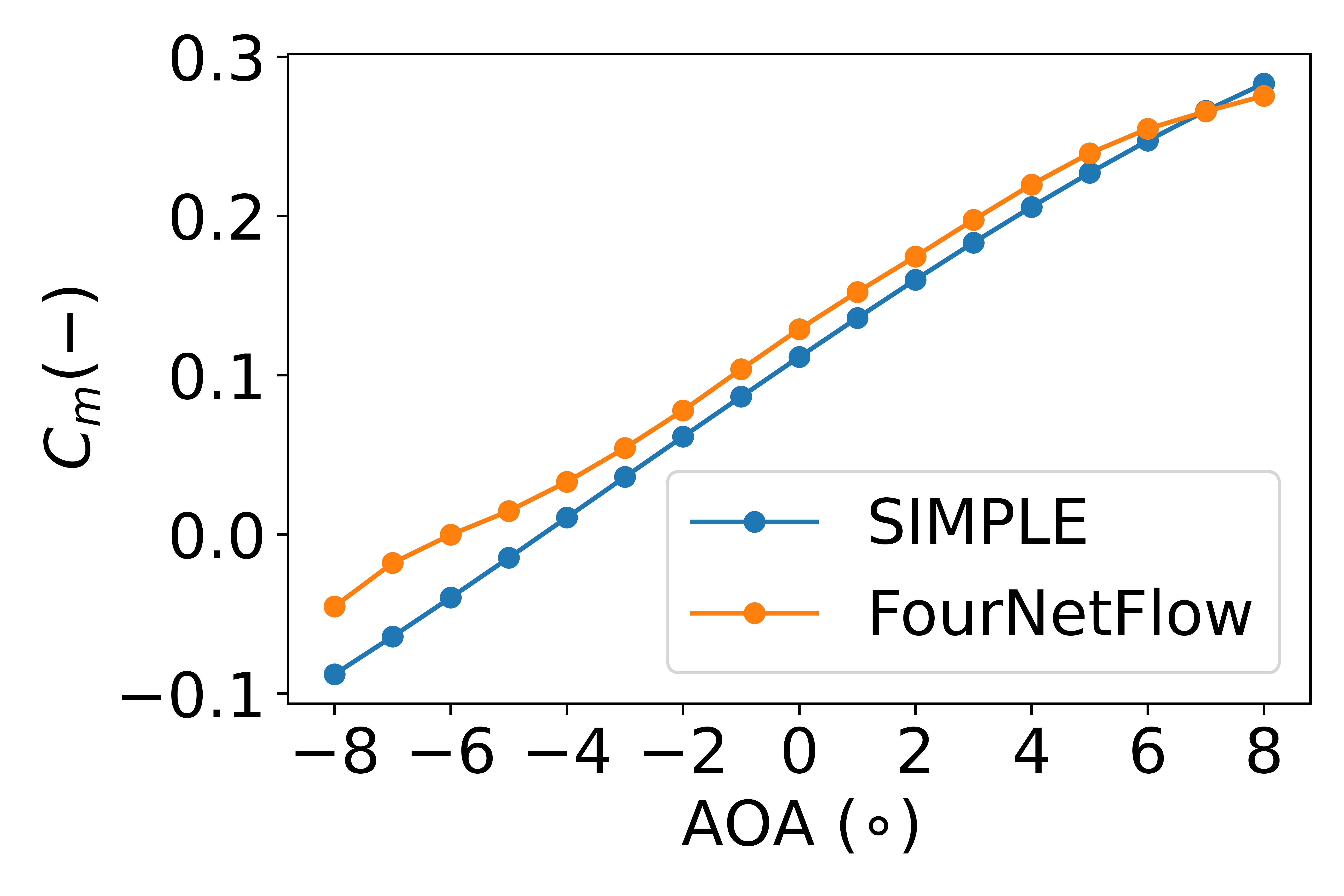}}
\quad
\subfloat[]{%
\includegraphics[totalheight=1.2in]{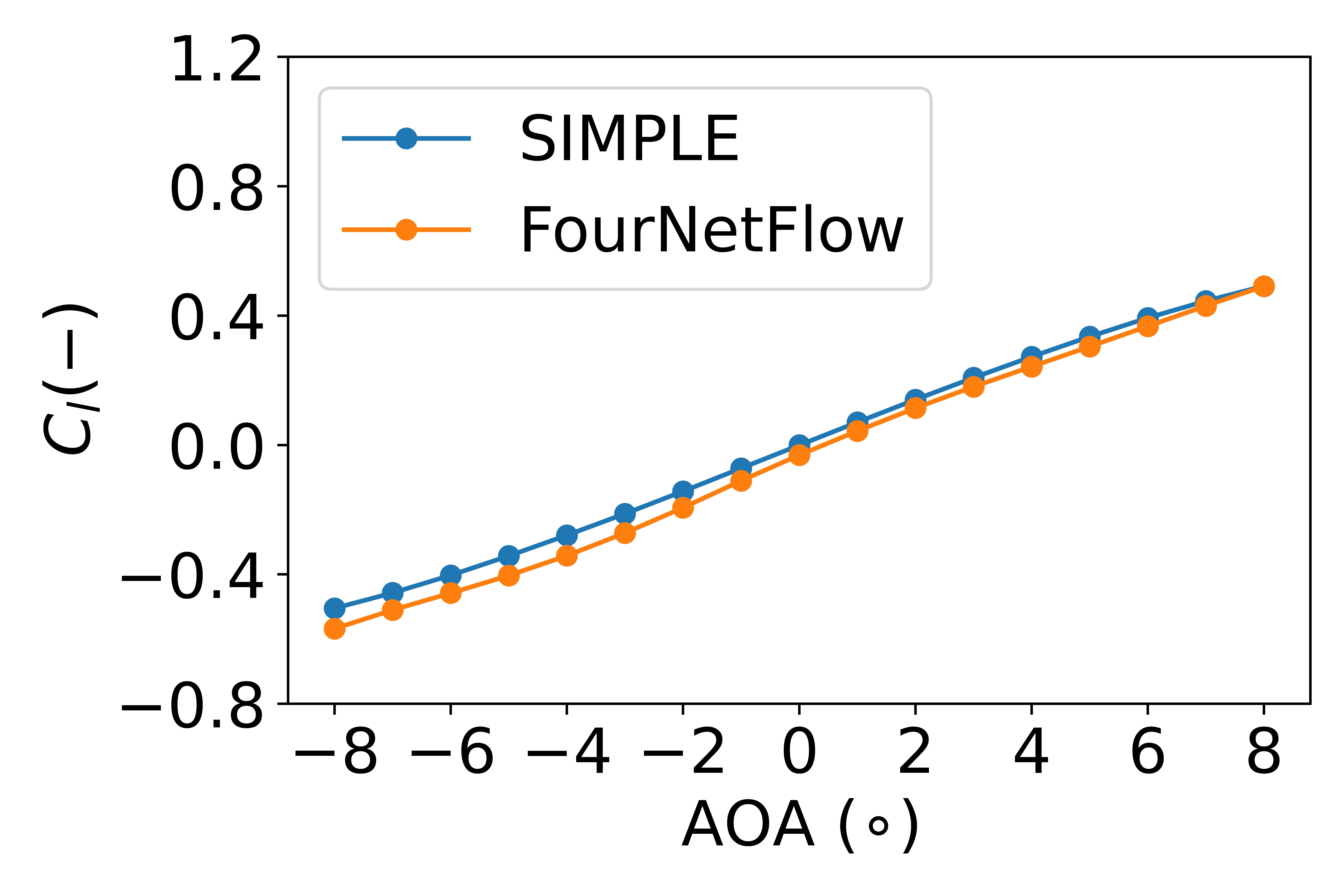}}
\subfloat[]{%
\includegraphics[totalheight=1.2in]{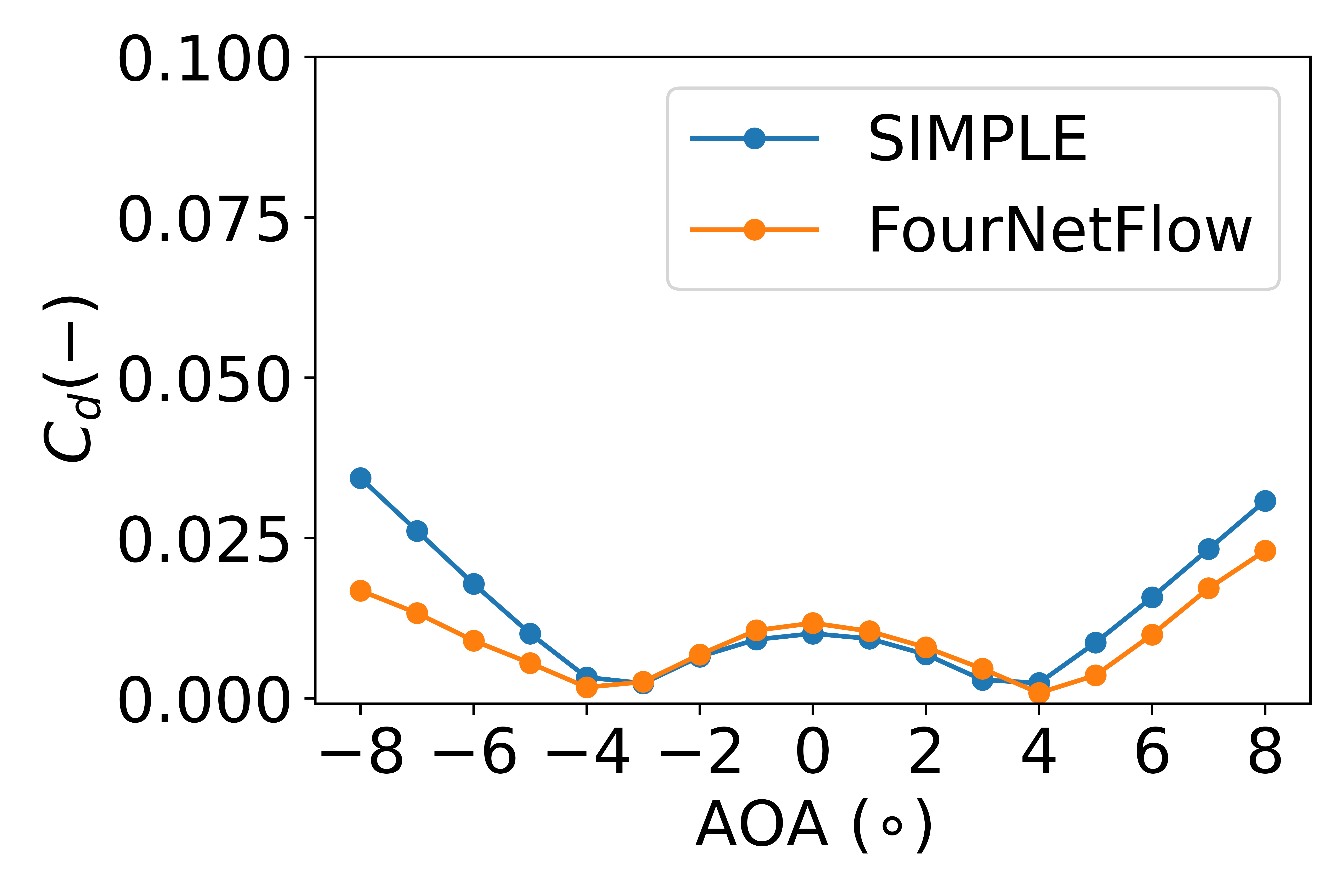}}
\subfloat[]{%
\includegraphics[totalheight=1.2in]{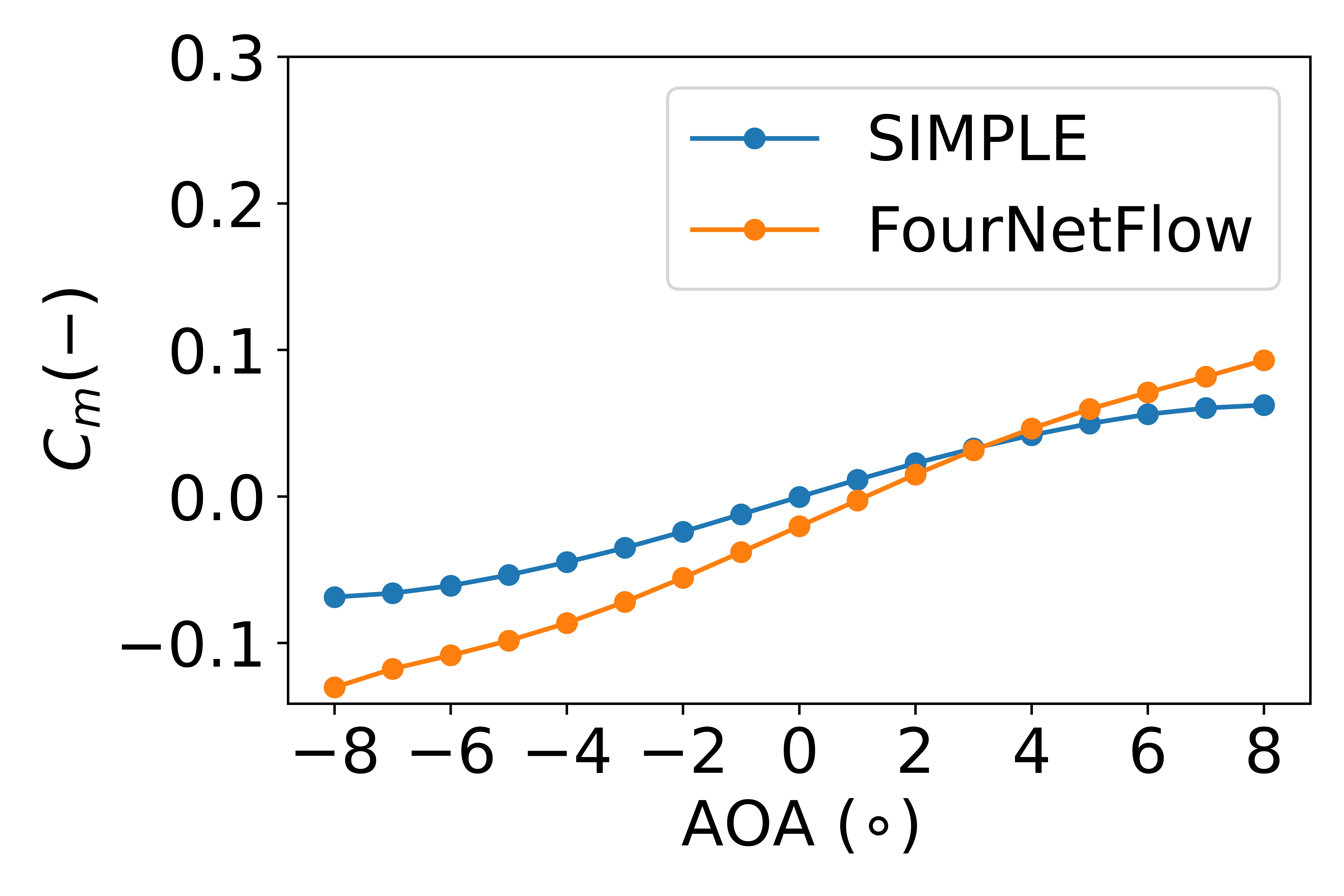}}
\caption{Comparison of integral quantities predicted by two methods. Panel (a)-(c) respectively shows the $C_l$, $C_d$, $C_m$ of NACA0012. Panel (d)-(f) shows these of RAE2822. Panel (g)-(i) shows these of OVAL10.}
\label{fig:fig5}
\end{figure}

\subsection{Super-resolution}
\begin{figure}
\centering
\includegraphics[totalheight=4in]{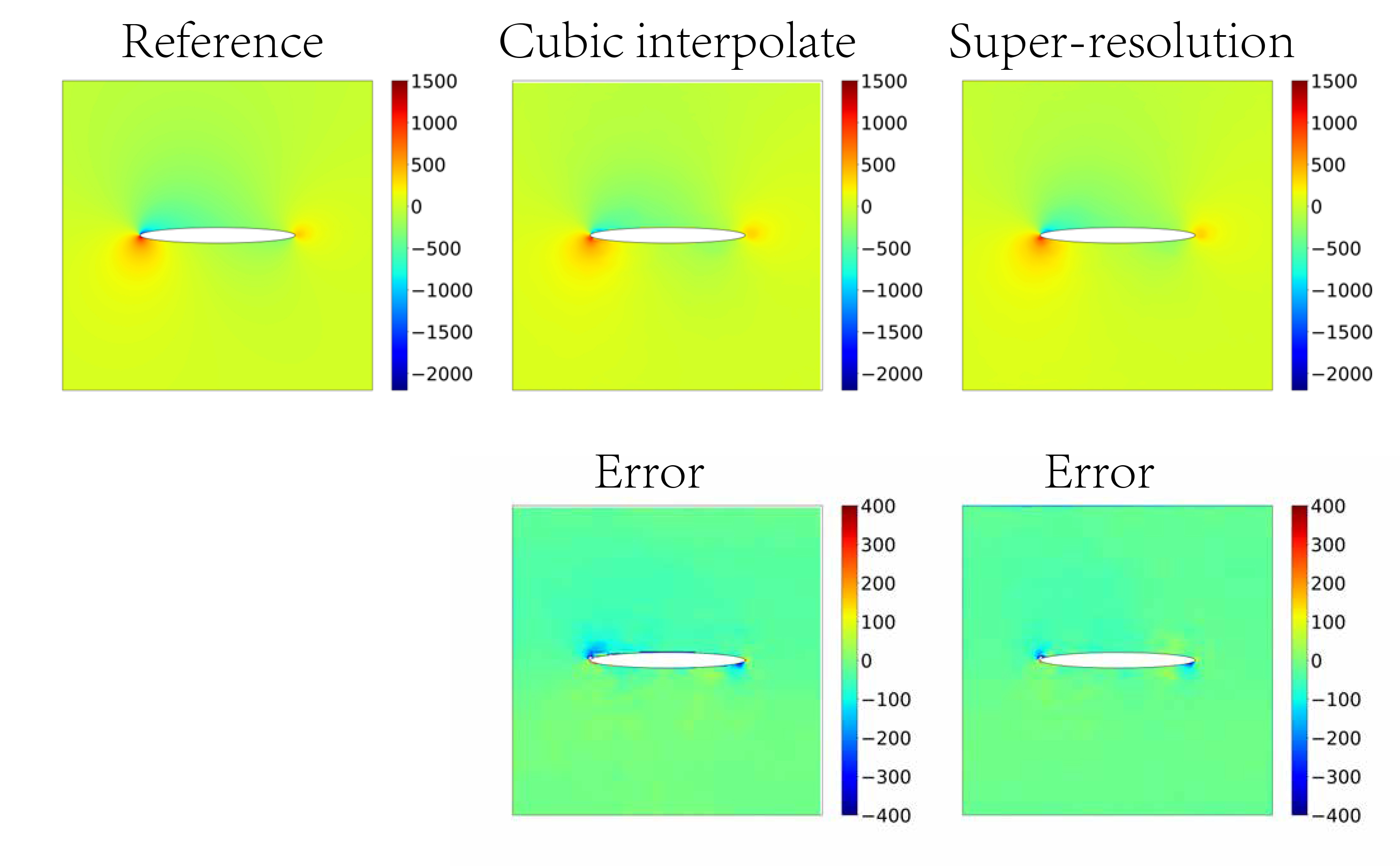}
\caption{The colour contours of pressure $p$ at resolution of 1024x1024. The \emph{Reference} is obtained from SIMPLE results. The \emph{Cubic interpolate} is interpolated from the 128x128 pressure field predicted by the FourNetFlows. The \emph{Super-resolution} is directly predicted by the FourNetFlows at resolution of 1024x1024. The \emph{Error} denotes the difference between the results of these two methods and the \emph{Reference}.}
\label{fig:fig6}
\end{figure}

To demonstrate zero-shot super-resolution of coarse flow data, we consider the application of the FourNetFlows model that trained by samples at resolution of $128\times128$ to directly predict the flow field of OVAL10 at $1024\times1024$. We compare the results with a simple cubic interpolation \cite{30}, which is a traditional approach. Figure \ref{fig:fig6} shows the comparison between the cubic interpolation and super-resolution. We note that the interpolation routine exhibits qualitative agreement with the reference flow field, but with a larger error over around the surface of the oval relative to the super-resolution. This shows that our model does not just interpolate from the training set to generate a similar result, but actually learns the underlying laws.

\subsection{Computational Cost of FourNetFlows }
In comparing the computational cost of physical fields generation between FourNetFlows and SIMPLE, we encounter a rather difficult problem of comparing a solver using a CPU (in the case of the SIMPLE) and a predictor that is run on a few GPUs (FourNetFlows). We don't have a solution to this problem for the time being, so we compare the computation times directly.
\begin{table}
 \caption{The computational time of SIMPLE and FourNetFlows}
  \centering
  \begin{tabular}{lllll}
    \toprule
    \cmidrule(r){1-2}
    Algorithm     & Numbers of cases     & Resolution  & Average time (s) \\
    \midrule
    SIMPLE & 1280  & $128\times128$    & 60.00 \\
    FourNetFlows & 1 & $128\times128$  & 1.04 \\
    FourNetFlows & 1280 & $128\times128$  & 0.005 \\
    FourNetFlows & 1 & $1024\times1024$  & 1.05 \\
    FourNetFlows & 1280 & $1024\times1024$  & 0.006 \\
    \bottomrule
  \end{tabular}
  \label{tab:table1}
\end{table}
Table \ref{tab:table1} presents the computational time of SIMPLE and FourNetFlows. The FourNetFlows neural has an inference time of only 0.005 s when inference for a batch of cases compared to the 60 s of the SIMPLE used to solve Navier-Stokes. And when the grids increase, the computation time keep an constant. It is almost unbelievable for conventional methods.

\section{Conclusion}
\label{sec:conclusion}
We developed an efficient model, FourNetFlows, for steady flow around airfoils based on the FNO and applied to predict steady flow fields of NACA0012, RAE2822 and OVAL10. FourNetFlows's accuracy is impressive compare to traditional methods. Its ability to directly generate super-resolution solution from a family of low-resolution pairs and generalize from a series of solutions of airfoil flows to the solution of oval flows has important implications for our development for a novel solver based neural networks. 

FourNetFlows's predictions are four to five orders of magnitude faster than SIMPLE. This has two important implications. First, FourNetFlows is suitable for rapidly testing hypotheses about mechanisms of physical quantities and their predictability. Second, the unprecedented speed in forecasts of flow fields and aerodynamics has potentially massive benefits for airfoil designing industry. The highly accurate and efficient model maybe substitute for the expensive CFD in optimization process. 

Predictions of the FourNetFlows is agreement with the results of the SIMPLE that is used to generate our training sets. But there is still a gap between our model's predictions and experimental results on small quantities like $C_d$. Maybe it is resulted from our dataset itself is not precise enough. It is well known that data-driven methods rely on the quality and quantity of data and generating a few training samples by CFD methods can be already expensive. Therefore, we compromise on the quality of data when generating training sets by SIMPLE. In terms of results, we believe that the accuracy of our model predictions will improve further when the accuracy of the dataset increases. Although we are using samples of airfoil flows to train this model, we know from the training framework that this model can be applied to any steady flow if you can get enough training sets. We look forward to more funding and computational resources to make our model more precise and more universal.

\section{Future work}

FourNetFlows is a purely data-driven DL model. The physical systems of flows are governed by the laws of nature, some of which are well-understood. DL models that obey the laws of physics are more likely to be robust. An emerging field in ML/DL applications in the sciences is Physics-informed Machine Learning. Future versions of our model will incorporate physical laws. 

\section*{Acknowledgments}
We acknowledge the High-performance Computing Platform of Peking University for providing computational resources. This work was supported by the Basic Research Program [JCKY2018204b054].
\newpage

\newpage
\appendix

\section{Data generation \& Preprocessing}
\label{appendix a}
We refer to the method of data generation in Thuerey et al. work \cite{26}. First, we obtain 1505 different airfoil shapes from the UIUC database. Then, we use the open source mesh tool, \emph{gmsh}, to generate a body-fitted triangle mesh with refinement near the airfoil. Finally, the RANS simulations make use of the widely used SA one equation turbulence model, and solutions are calculated with \emph{OpenFOAM}. And some details in our work are different. We use a lager domain to reduce the boundaries' negative impact on the solutions around the airfoil. We consider a space of solutions with a range of Reynolds numbers $Re = [1,10]$ millions, incompressible flow. And angles of attack is limited in the range of $\pm 10$ deg to ensure that it is a steady state problem. We increase the number of iterations to make the residual of each case convergence. These issues were not considered in Thuerey et al. codes. Some of their samples showed significant unsteady effects, and the residual of some cases still kept a relatively high level.

We resample a smaller refion of $2\times2$ units around the airfoil with a Cartesian $128^2$ grid as inputs of FourNetFlows. The resampleing is performed with a linear weighted interpolation of cell-centered values with a spacing of 1/64 units in \emph{OpenFOAM}. Interpolation inevitably introduces errors. We ensure the scale of computational grids is smaller than 1/64 units to alleviate the issue.

Since we assume it is an incompressible problem, the solutions globally depend on freestream conditions and the airfoil shape, we naturally plan to encode these conditions into some tensors with a size of $128\times128$. The major problem is to distinguish points outside the airfoil which should be given speed, and points inside the airfoil which should be given solid wall conditions. We utilize the \emph{PNpoly} algorithm to solve the issue. \emph{PNpoly} runs a semi-infinite ray horizontally (increasing X, fixing Y) from the test point and count how many edges of the polygon it crosses. If it passes through an even number, the point is outside the polygon, otherwise it is inside the polygon. With the algorithm, we obtain two input tensors, $x$ and $y$ velocity component. Both velocity channels are initialized to the maximum of freestream velocity with a zero velocity inside of the airfoil shape. We note that a additional input tensor that contains a $[0,1]$ mask for the airfoil shape will increase the accuracy. Therefore, we take mask, $x$ and $y$ velocity component as inputs of our model.

\begin{figure}
\centering
\includegraphics[totalheight=2in]{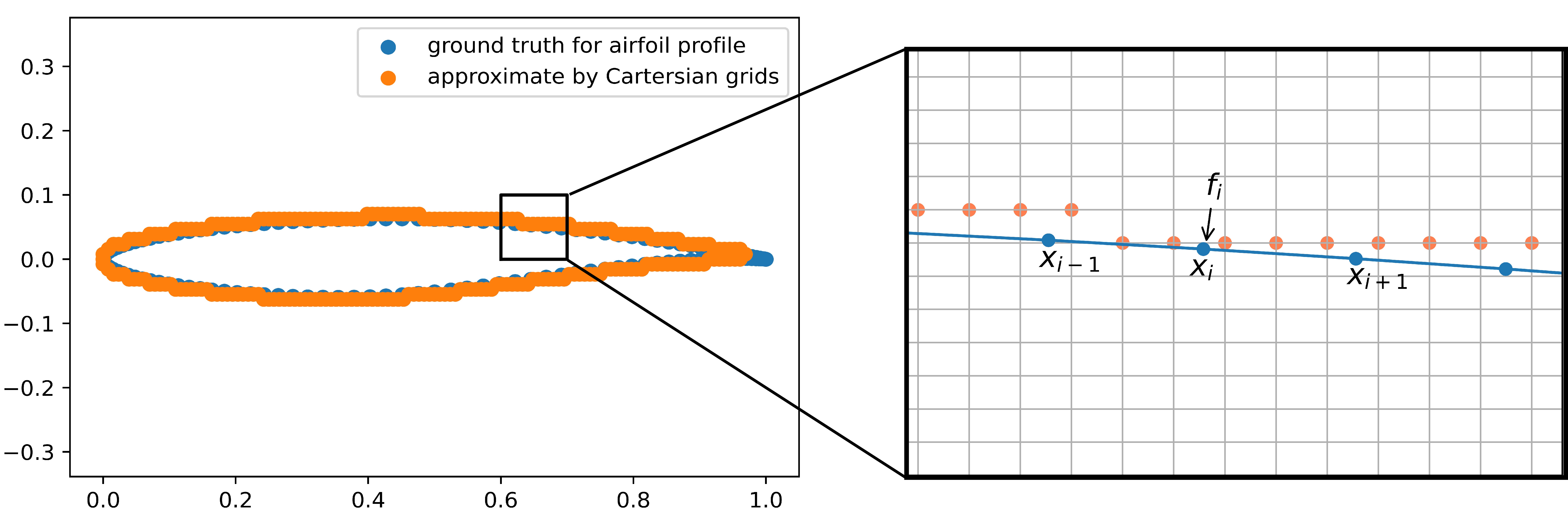}
\caption{Illustrative example of the approximate profile and ground truth of \emph{RAE2822}. Right panel is a partial enlargement of the left panel, illustrating how to calculate the external force of the \emph{i} th point.}
\label{fig:appendix1}
\end{figure}

As mentioned above, FourNetFlows outputs the pressure and velocity distribution around the airfoil on a Cartesian $128^2$ grid. 
It is difficult to describe the shape of the airfoil accurately via Cartesian grid due to curves have to be expressed approximately in zig-zag form.
Figure \ref{fig:appendix1} shows the approximate profile of the airfoil and the ground truth. Coral points denote the approximate airfoil and blue points represent the exact airfoil profile. There obviously are some differences between the two representations, which is a more severe problem if it is expressed at lower resolution Cartesian grids. Therefore, the pressure on the surface of airfoils output by FourNetFlows is actually the pressure on the surface of approximate airfoils. Since the approximate airfoil is a zig-zag representation, it is incorrect to directly do a curvilinear integration under the approximate airfoil. To compute integral quantities, we firstly need to obtain the pressure on the surface of exact airfoils. 
We take the following assumption: the pressure of each point on the exact airfoil is equal to the pressure of the nearest approximate airfoil point in the normal direction. 
This approximation is reasonable since the scale of grids close to the airfoil wall is small enough and the pressure gradient within the first layer is almost zero. After obtaining the pressure on the exact airfoil, we compute integral quantities by a approximate curvilinear integration. We take the three points on the exact airfoil shown on the right panel of Fig \ref{fig:appendix1} as an example of how we obtain integral quantities.
\begin{equation}
\begin{aligned}
    \textbf{f}_i &= \frac{1}{2}p_i*|\textbf{X}_{i-1}-\textbf{X}_i|*\textbf{n}_{i-1,i}+\frac{1}{2}p_i*|\textbf{X}_{i}-\textbf{X}_{i+1}|*\textbf{n}_{i,i+1}\\
    C_l &= \frac{\sum \textbf{f}_{i,y}}{\frac{1}{2} \rho U_{\infty}^2 S}, ~~~~~C_d = \frac{\sum \textbf{f}_{i,x}}{\frac{1}{2} \rho U_{\infty}^2 S }, ~~~~~C_m = \frac{\sum (-\textbf{f}_{i,x}*\textbf{X}_{i,y} + \textbf{f}_{i,y}*\textbf{X}_{i,x})}{\frac{1}{2} \rho U_{\infty}^2 S L}
\end{aligned}
\end{equation}
where $\textbf{X}_i$ is the coordinate of the \emph{i} th point of the exact airfoil. $\textbf{f}_i$ is the external force at the \emph{i} th point. $p_i$ is the pressure at the \emph{i} th point, which is inferred from the pressure of the approximate airfoil. $n_{i-1,i}$ is the normal vector of the line connecting the \emph{i-1} th point and the \emph{i} th point. $\rho$ is the density of fluid. $U_{\infty}$ is the freestream velocity. $S$ is the reference area and $L$ is the reference length.

\end{document}